\documentclass[prb,showpacs]{revtex4}
\usepackage{graphicx,psfrag,amssymb,amsmath,color}
\begin{document}
\title{Magnetotransport and thermoelectricity in disordered graphene}
\author{Bal\'azs D\'ora}
\email{dora@kapica.phy.bme.hu}
\affiliation{Max-Planck-Institut f\"ur Physik Komplexer Systeme, N\"othnitzer Str. 38, 01187 Dresden, Germany}
\author{Peter Thalmeier}
\affiliation{Max-Planck-Institut f\"ur Chemische Physik fester Stoffe, 01187 Dresden, Germany}

\date{\today}

\begin{abstract}
We have studied the electric and thermal response of two-dimensional Dirac-fermions in a quantizing magnetic field in 
the presence of localized disorder. The electric and heat current operators in the presence of magnetic field 
are 
derived. The self-energy due to impurities is calculated self-consistently, and depends strongly on the frequency and 
field strength, resulting in 
asymmetric peaks in the density of states  at the Landau level energies, and small islands connecting them. The 
Shubnikov-de Haas oscillations remain 
periodic in $1/B$, in spite of the distinct quantization of quasiparticle orbits compared to normal metals.
The Seebeck coefficient depends strongly on the field strength and orientation.
For finite field and chemical potential, the Wiedemann-Franz law can be violated.
\end{abstract}


\maketitle

\section{Introduction}
Recent advances of nanotechnology have made the creation and investigation of two dimensional carbon, called graphene, 
possible\cite{berger,novoselov1,novoselov3,bostwick}. 
It is a monolayer of carbon atoms packed densely in a honeycomb structure. In spite of being few atom thick, these 
systems were found to be stable and ready for exploration. 
One of the most intriguing property of graphene is, that its charge carriers are well described by the relativistic 
Dirac's equation, and are two-dimensional Dirac fermions\cite{zhou}. This opens the possibility of investigating 
"relativistic" 
phenomena at a speed of $\sim 10^6$~m/s (the Fermi velocity of graphene), 1/300th the speed of light. 
The linear, Dirac-like spectrum causes the density of states to increase linearly with energy, which is to be 
contrasted the constant density of states of normal metals. Due to this peculiar property, the response of graphene to 
external probes is expected to be unusual. 
This manifests itself in the anomalous integer quantum Hall effect\cite{novoselov2}, which occurs at half-integer 
filling factors, and in the presence of universal minimal value of the conductance.
The dependence of the thermal conductivity on applied magnetic field has been measured in highly oriented pyrolytic 
graphite\cite{ocana,ulrich}.

Dirac-fermions show up in other systems, at least from the theoretical side. They characterize the low energy 
properties of orbital antiferromagnets, a density wave system with a gap of d-wave symmetry\cite{Ners1,Ners2}. A 
similar 
model has been proposed for the pseudogap phase of high $T_c$ cuprate superconductors, known as d-density wave, with 
peculiar electronic 
properties\cite{nayak}. A similar system was also mentioned in the context of heavy fermion material URu$_2$Si$_2$, 
which 
shows a clear phase transition at 17~K without any obvious long range order, detectable by X-ray or NMR experiments. 
Its low temperature phase was attributed to another spin density wave  with a d-wave gap\cite{IO,roma}
Experimentally, the aforementioned materials possess unusual electric and thermal responses as a function of 
temperature 
and magnetic field\cite{behnia1,behnia2}.

Therefore, the interest in studying the transport properties of two-dimensional Dirac-fermions is not surprising. 
Gusynin, Sharapov and coworkers have studied 
exhaustively\cite{sharapov1,sharapov2,sharapov3,sharapov4,sharapov5,sharapov6,sharapov7} the electric and thermal 
response of two-dimensional systems with linear energy spectrum, with special emphasis on the Wiedemann-Franz law and 
magnetic oscillations. However, their self energy due to scattering from impurities was not determined in a 
self-consistent manner, but rather they assumed a constant, energy, magnetic field and temperature independent 
scattering rate. Moreover, they completely neglected the real part of the self energy, responsible for 
the shift of energy levels. Nevertheless, they derived beautiful analytical formulas for the various transport 
coefficients, which, although suffering from the above limitations, turned out to be useful in explaining 
experiments\cite{novoselov2}.

Impurity scattering can be taken into account in the presence of quantizing magnetic field in the usual self-consistent 
way\cite{klasszikus}. This program has been carried out, among many others\cite{peres1}, by Peres et 
al.\cite{peresalap}. In their 
work, the full self-consistent Born approximation was used before taking the strength of the impurity potential to 
infinity. 
They studied the frequency dependence of the electric conductivity for various fields, but never entered into the realm 
of 
thermal transport. 
Parallel studies have also been performed in the limit of weak-scatterers\cite{ando1,ando2}.

In this paper, we extend the work of Refs. \onlinecite{sharapov1,sharapov2,sharapov3,sharapov4}, and determine 
self-consistently the energy and magnetic field dependent self-energies and study the Seebeck coefficient as well, and 
also generalize Ref. 
\onlinecite{peresalap} to include thermoelectricity. We study Dirac-fermions in a Landau quantizing magnetic field 
($B$) in 
the 
presence of scatterers, allowing for arbitrary field orientations. 
In a way, our study here bridges between the efforts of the previous groups. After the introduction of the general 
formalism, 
we determine the electric and heat current operators, essential for further steps. By introducing impurities in the 
system, we can study the quasiparticle  density of states, the electric and heat conductivity, the Seebeck coefficient 
and the Wiedemann-Franz law as a function of magnetic field strength and orientation and temperature. For high fields, 
the discrete nature of the Landau levels is revealed in the density of states in the form of asymmetric peaks at 
Landau level energies (far from being Lorentzians), which smoothen with decreasing field. 
Shubnikov-de Haas oscillation are visible in all transport coefficients, periodic in $1/B$, similarly to 
normal metals\cite{abrikosov}. The angular dependent conductivity oscillations become more pronounced with 
increasing field. 
The chemical potential dependence of the conductivity resembles closely the experimental findings\cite{novoselov2}.
The Seebeck coefficient depends strongly on the applied field and temperature.

\section{Landau quantization, electric and heat current}

The Hamiltonian of non-interacting quasiparticles living on a single graphene sheet is given 
by\cite{semenoff,gonzalez,peresalap}:
\begin{gather}
H_0=-v_F\sum_{j=x,y}\sigma_j\left(-i\partial_j+eA_j(\bf r)\right),
\label{hamalap}
\end{gather}
where $\sigma_j$'s are the Pauli matrices, and stand for Bloch states residing on the two different sublattices of the 
bipartite hexagonal lattice of graphene\cite{peresalap,sharapov3}.
Strictly speaking, the  Hamiltonian above describes quasiparticles around the $K$ points of the Brillouin zone, where 
the spectrum vanishes. 
The 
vector potential for a constant, arbitrarily oriented  magnetic 
field reads as ${\bf A(r)}=(-By\cos\theta,0,B(y\sin\theta\cos\phi-x\sin\theta\sin\phi))$, where $\theta$ is the 
angle the magnetic field makes from the $z$ axis, and $\phi$ is the in-plane polar-angle measured from the $x$-axis.
We have dropped the Zeeman term, its energy would be negligible with respect to energy of the Landau levels, 
Eq. 
\eqref{landauenergia}, using $v_F\approx 10^6$~m/s, characteristic to graphene. Eq. \eqref{hamalap} applies for both 
spin directions.

In the absence of magnetic field, the energy spectrum of the system is given by 
\begin{equation}
E({\bf k})=\pm v_F|\bf k|.
\end{equation}
This describes massless relativistic fermions with spectrum consisting of two cones, touching each other at the 
endpoints.
From this, the density of states per spin follows as
\begin{equation}
\rho(\omega)=\frac 1\pi \sum_{\bf k}\delta(\omega-E({\bf k}))=\frac 1\pi 
\frac{A_c}{2\pi}\int\limits_0^{k_c}kdk\delta(\omega\pm v_F k)=\frac{2|\omega|}{D^2},
\end{equation}
where $k_c$ is the cutoff, $D=v_Fk_c$ is the bandwidth, and $A_c=4\pi/k_c^2$ is the area of the hexagonal unit cell.

In the presence of magnetic field, the eigenvalue problem of this Hamiltonian ($H_0\Psi=E\Psi$) can readily be 
solved\cite{peresalap}. 
For the zero energy mode (E=0), the 
eigenfunction is obtained as
\begin{gather}
\Psi_k({\bf r})=\frac{e^{ikx}}{\sqrt{L}}\left(\begin{array}{c}
0 \\
\phi_0(y)
\end{array}\right),
\end{gather} 
and the two components of the spinor describe the two bands. The energy of the other modes read as
\begin{gather}
E(n,\alpha)=\alpha\omega_c\sqrt{n+1}
\label{landauenergia}
\end{gather}
with $\alpha=\pm 1$, $\omega_c=v_F\sqrt{2e|B\cos(\theta)|}$ is the cyclotron frequency, $n=0$, 1, 2,\dots. 
Only the perpendicular component of the field enters into these expressions, and by tilting the field away from the 
perpendicular direction corresponds to a smaller effective field.
The sum over integer $n$'s is cut off  
at $N$ given by $N+1=(D/\omega_c)^2$, which means that we consider $2N+3$ Landau levels altogether. For later 
convenience, we define a magnetic field $B_0$, 
whose cyclotron frequency is equal to the bandwidth ($\omega_c=D$).

\begin{figure}[h!]
\psfrag{n0}[l][r][1][0]{$E=0$}
{\includegraphics[width=5cm,height=5cm]{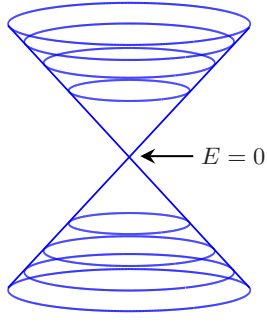}} 
\caption{The structure of the Landau levels is visualized schematically for the first few levels.\label{landaul}}
\end{figure}

The corresponding wave function is
\begin{gather}
\Psi_{n,k,\alpha}({\bf r})=\frac{e^{ikx}}{\sqrt{L}}\left(\begin{array}{c}
\phi_n(y-kl_B^2) \\
\alpha\phi_{n+1}(y-kl_B^2)
\end{array}\right)
\end{gather}  
with cyclotron length $l_b=1\sqrt{eB}$. Here $\phi_n(x)$ is the $n$th eigenfunction of the usual one-dimensional 
harmonic oscillator. The electron-field operator can be built up from these functions as
\begin{gather}
\Psi({\bf r})=\sum_k\left[ \Psi_k({\bf r})c_{k}+\sum_{n,\alpha}\Psi_{n,k,\alpha}c_{k,n,\alpha}\right].
\end{gather}
The Green's functions of these new operators do not depend on $k$, and read as
\begin{gather}
G_0(i\omega_n,k)=\frac{1}{i\omega_n},\\
G_0(i\omega_n,k,n,\alpha)=\frac{1}{i\omega_n-E(n,\alpha)}
\end{gather}
for $c_k$ and $c_{k,n,\alpha}$, respectively, and $\omega_n$ is the fermionic Matsubara frequency.

With the use of these, we can determine the electric and heat current operators of the system.
Following Mahan\cite{mahan}, we define the polarization operator as
\begin{equation}
{\bf P}=\frac 12\int d{\bf r}\left({\bf r}\rho({\bf r})+\rho({\bf r}){\bf r}\right)
\end{equation}
with $\rho({\bf r})=\Psi^+({\bf r})\Psi({\bf r})$ giving the charge density, and the symmetric combination ensures 
hermiticity.
The total current is its time derivative, which follows as
\begin{equation}
{\bf J}=\partial_t{\bf P}=i[H,{\bf P}].
\end{equation}
By performing the necessary steps, after straightforward calculations this yields\cite{peresalap}
\begin{gather}
J_x=v_Fe\sum_{p,\alpha}\left[ \frac{1}{\sqrt 2}(c^+_{p}c_{p,0,\alpha}+c^+_{p,0,\alpha}c_{p})+\sum_{n,\lambda}
\frac \lambda 2
(c^+_{p,n+1,\alpha}c_{p,n,\lambda}+c^+_{p,n,\lambda}c_{p,n+1,\alpha})\right],\\
J_y=iv_Fe\sum_{p,\alpha}\left[ \frac{1}{\sqrt 2}(c^+_{p}c_{p,0,\alpha}-c^+_{p,0,\alpha}c_{p})+\sum_{n,\lambda}
\frac{\lambda}{2}
(c^+_{p,n,\lambda}c_{p,n+1,\alpha}-c^+_{p,n+1,\alpha}c_{p,n,\lambda})\right].
\end{gather} 
The heat current operator for the pure system can be determined similarly.
In analogy with polarization, one defines the energy position operator\cite{jonson} as
\begin{equation}
{\bf R}^E=\frac 12\int d{\bf r}\left({\bf r}H({\bf r})+H({\bf r}){\bf r}\right),
\end{equation}
and the total Hamiltonian is $H=\int d{\bf r}H(\bf r)$. Using this, one deduces the energy current from
\begin{equation}
{\bf J}^E=\partial_t{\bf R}^E.
\end{equation}
This leads to
\begin{gather}
J^E_x=\frac{v_Fe}{2}\sum_{p,\alpha}\left[ \frac{E(0,\alpha)}{\sqrt 
2}(c^+_{p}c_{p,0,\alpha}+c^+_{p,0,\alpha}c_{p})+\sum_{n,\lambda}
\frac \lambda 2 (E(n+1,\alpha)+E(n,\lambda)) 
(c^+_{p,n+1,\alpha}c_{p,n,\lambda}+c^+_{p,n,\lambda}c_{p,n+1,\alpha})\right],\\
J^E_y=\frac{iv_Fe}{2}\sum_{p,\alpha}\left[ \frac{E(0,\alpha)}{\sqrt 
2}(c^+_{p}c_{p,0,\alpha}-c^+_{p,0,\alpha}c_{p})+\sum_{n,\lambda}
\frac{\lambda}{2}(E(n+1,\alpha)+E(n,\lambda))
(c^+_{p,n,\lambda}c_{p,n+1,\alpha}-c^+_{p,n+1,\alpha}c_{p,n,\lambda})\right].
\end{gather}
These follow naturally from the electric current operator, after multiplying each term with the corresponding 
mode energy.
Note, that the energy of the state labeled solely by $(p)$ is zero, it belongs to the state situated at the meeting 
point of the two cones. 
Finally the heat current operator is related to the energy current by the simple formula: ${\bf J}^Q={\bf J}^E-\mu{\bf 
J}$, $\mu$ is the chemical potential. So far we 
have 
considered the particle-hole symmetric case with $\mu=0$, but we can easily use a finite chemical potential to break 
this symmetry, and introduce finite Seebeck coefficient.

\section{Impurity scattering in the presence of magnetic field}

In the presence of impurities, an extra term is added 
to the Hamiltonian: \begin{equation}
H_{imp}=V\sum_{i=1}^{N_i}\delta({\bf r-r}_i).
\label{hamimp}
\end{equation}
As a result, the explicit form of the previous operators might change.  However, using Eq. \eqref{hamimp},
the electric current remains unchanged, but the heat current changes due to the non-commutativity of the impurity 
Hamiltonian and the energy position operator\cite{mahan}. As a result, impurities need to be taken into account not 
only in the 
calculation of the self-energy, but also in the form of the operators, and one has to use the same level of 
approximation for both. 

However, to avoid this difficulty, one can replace the energy terms in ${\bf J}^E$ by the Matsubara 
frequency\cite{jonson}, since 
from  the poles of the Green's function, this will pick the appropriate energy. This replacement works perfectly in the 
case of impurities as well, when quasiparticle excitations possess finite lifetime.

Since graphene is two-dimensional, positional long range order (i.e. lattice formation) is impossible at finite 
temperatures, since thermal fluctuations will destroy it. This is why the introduction of defect is natural in this 
system. To mimic disorder, we have chosen to spread vacancies in the honeycomb lattice, which can be modeled by taking 
the impurity strength ($V$) to infinity.

To take scattering into account, we have to determine the explicit form of $H_{imp}$ in the Landau basis. Then the 
standard non-crossing approximation can be used\cite{klasszikus}, which, in the case of graphene, is called the full 
self-consistent 
Born approximation due to the neglect of crossing diagrams\cite{ando1,ando2,peresalap}. Averaging over impurity 
positions is performed in the 
standard way. By letting the impurity strength $V\rightarrow \infty$, 
which would correspond to the unitary scattering limit in unconventional superconductors\cite{impurd-wave}, we arrive 
to the following 
set of equations\cite{peresalap}:
\begin{gather}
G(i\omega_n,k,n,\alpha)=\frac{1}{i\omega_n-E(n,\alpha)-\Sigma_1(i\omega_n)},\\
G(i\omega_n,k)=\frac{1}{i\omega_n-\Sigma_2(i\omega_n)},
\end{gather}
where
\begin{gather}
\Sigma_1(i\omega_n)=-\frac{2n_i}{g_c[G(i\omega_n,k)+\sum_{n,\alpha}G(i\omega_n,k,n,\alpha)]},\label{self1}\\
\Sigma_2(i\omega_n)=-\frac{2n_i}{g_c[2G(i\omega_n,k)+\sum_{n,\alpha}G(i\omega_n,k,n,\alpha])},\label{self2}
\end{gather}
where $g_c=1/(N+1)$ is the degeneracy of a Landau level per unit cell, $n_i$ is the impurity 
concentration. The summation over Landau levels can be performed to yield
\begin{equation}
\sum_{n,\alpha}G(i\omega_n,k,n,\alpha)=2\frac{z}{\omega_c}\left\{\Psi(1-z^2)-\Psi(N+2-z^2)\right\},
\end{equation}
where $z=(i\omega_n-\Sigma_1(i\omega_n))/\omega_c$, $\Psi(z)$ is the digamma function.
The self-consistency equations can further be simplified, and after analytic continuation to real frequencies 
($i\omega_n\rightarrow 
\omega+i0^+$), we can read off
\begin{equation}
\Sigma_1(\omega)=\Sigma_2(\omega)\frac{2n_i(\omega-\Sigma_2(\omega))}{g_c\Sigma_2(\omega)+2n_i(\omega-\Sigma_2(\omega))}.
\end{equation}
At zero frequency,
this simplifies to
\begin{equation}
\Sigma_2(0)=\Sigma_1(0)\left(1-\frac{1}{2n_i(N+1)}\right).
\label{causal}
\end{equation}
The imaginary part of the self energy is always negative to ensure causality. This means that the last term in 
parenthesis on the right hand side must always be positive to assure the same sign of the imaginary parts of the self 
energies. This translates into 
\begin{equation}
n_i\geqslant\frac{1}{2(N+1)}.
\label{hatar}
\end{equation}
For each impurity concentration, there is a certain magnetic field strength (when $N=(1/2n_i)-1$), above which our 
approximation breaks down. 
For higher field, the self energy at zero frequency needs to be zero to fulfill Eq. \eqref{causal} and causality. This 
means, that at a finite impurity concentration, we still have excitations in the system with infinite lifetime.
Further, we are going to show that this occurs not only on the zeroth Landau level, but on all Landau levels for field 
exceeding the critical one.
To improve on this, crossing diagrams need to be considered, which is beyond the scope of the present work. Hence we 
restrict our investigation to fields allowed by Eq. \eqref{hatar}. The larger the impurity concentration, the larger 
the magnetic field we can take into account. 

The quasiparticle density of states can be evaluated from the knowledge of the Green's function, and it reads as
\begin{equation}
\rho(\omega)=\frac{2n_i}{\pi}\textmd{Im}\frac{1}{\Sigma_1(\omega)}.
\end{equation}

Without impurities, the density of states consists of Dirac-delta peaks located at zero frequency and at $E(n,\alpha)$. 
By introducing impurities in the system, we expect the broadening and shift of these levels, and it can be determined 
from the 
solution of the 
self consistency equations.

\begin{figure}[h!]
\psfrag{x}[t][b][1][0]{$\omega/D$}
\psfrag{y}[b][t][1][0]{$D\rho(\omega)$}
\centering{\includegraphics[width=7cm,height=7cm]{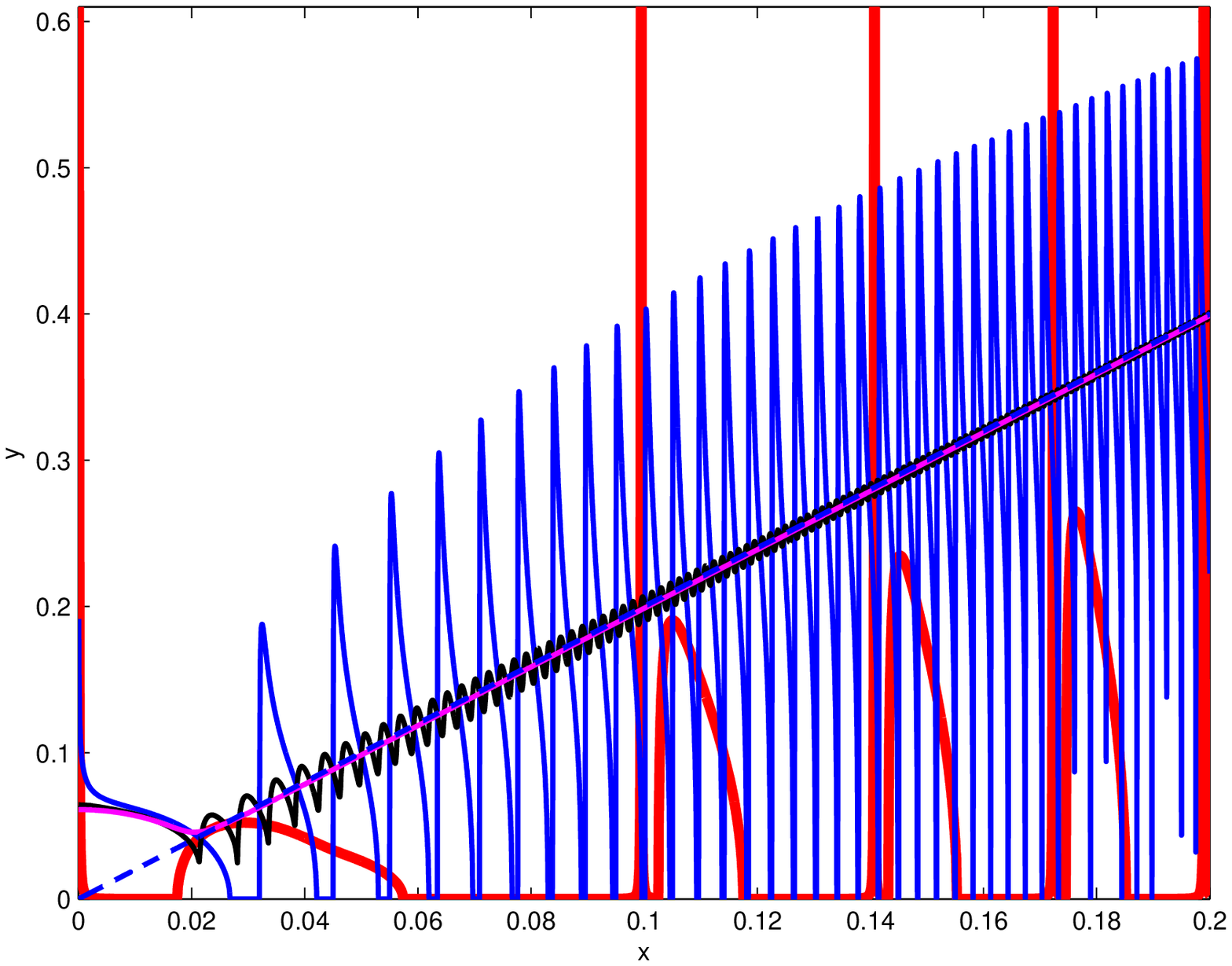}}
\hspace*{1cm}
\centering{\includegraphics[width=7cm,height=7cm]{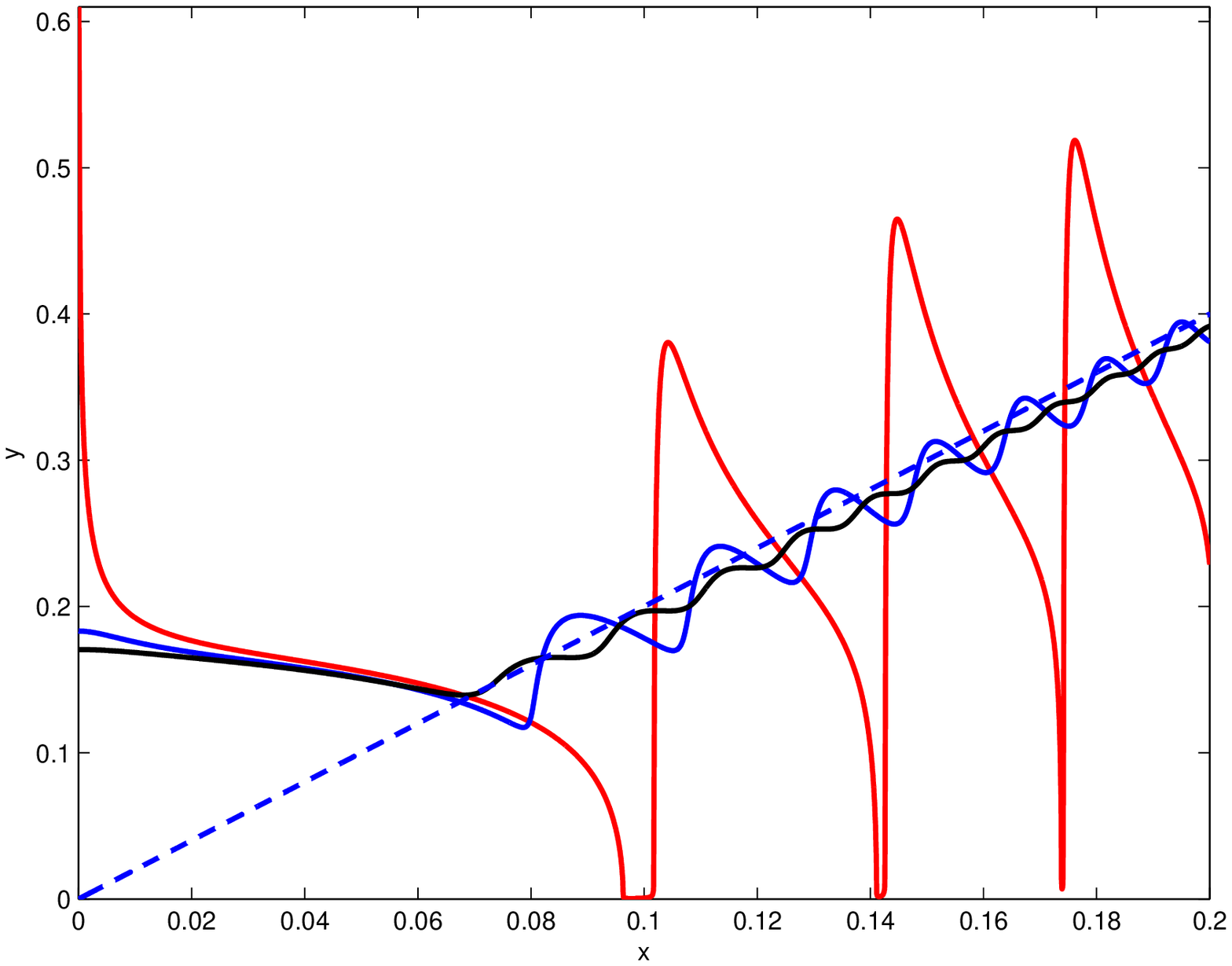}}
\caption{(Color online) The density of states is shown in the left panel for $n_i=0.001$ for $N=100$ (red), 1000 
(blue), 3000 
(black), 10000 (magenta). The vertical red lines stand for the Dirac-delta peaks for $N=100$. 
The right panel 
visualizes the $n_i=0.01$ case for $N=100$ (red), 200
(blue), 300
(black).
The clean case without 
magnetic field ($N=\infty$) is also plotted for comparison in both panels (blue 
dashed line).
\label{dosweak}}
\end{figure}

For large magnetic fields (small $N$), we can still solve the self-consistency equations Eqs. 
\eqref{self1}-\eqref{self2}, but we discover 
Dirac-delta peaks at the position of the levels and small islands between them (Fig. \ref{dosweak}, $N=100$). This 
signals that the 
non-crossing 
approximation is insufficient to provide these peaks with a finite broadening. As we decrease the field (increase $N$), 
the peaks and islands merge, and all excitations possess finite lifetime, but clean gaps are still observable 
between 
the levels. By further decreasing the field, the gaps disappear, the density of states becomes finite for all energies, 
and small successive bumps remain present due to Landau level formation, which tend to be smoothened by further 
decreasing the field. In this limit, the resulting density of states is very close to that in a d-wave 
superconductor\cite{hottacomment}, stemming from its linear frequency dependence in the pure case.

The broadening of the levels is not symmetric, more spectral weight is transferred to the lower energy part, which 
arises from the important energy dependence of the imaginary part of the self energies. 
Also the level position is modified in the presence of impurities due to the finite real part of the self energies, and  
this shift increases with the impurity concentration.

The numerical solution of Eqs. \eqref{self1}-\eqref{self2}, and the resulting density of states is shown in Fig. 
\ref{dosweak}. From this, one can conjecture, that a given $n_i$ and $N$ can qualitatively well describe different 
fields and concentrations, if their product ($n_iN$) is the same. These features, including the non-Lorentzian 
broadening of the Landau levels and the development of small islands between the levels should be observable 
experimentally by scanning tunneling microscopy, for example.

\section{Electric and thermal conductivities}

Using the spectral representation of the Green's functions, we can evaluate the corresponding conductivities after 
straightforward but lengthy calculations. 
These are related to the time ordered products of the form\cite{mahan}
\begin{equation}
\Pi^{AB}_{i,j}(i\omega)=-\int\limits_0^\beta d\tau e^{i\omega\tau}\langle T_\tau J^A_i(\tau)J^B_j(0)\rangle,
\end{equation}
where $A$ and $B$ denote the electric or heat current, $i$ and $j$ stand for the spatial component.
These can be expressed with the use of the following transport integrals\cite{abrikosov}:
\begin{gather}
L_n=\int\limits_{-\infty}^\infty \frac{d\epsilon}{4T}\frac{\sigma(\epsilon)}{\cosh^2((\epsilon-\mu)/2T)}
\left(\frac{\epsilon-\mu}{T}\right)^n,
\end{gather}
where
\begin{gather}
\sigma(\epsilon)=\omega_c^2\sum_\alpha\left[\frac{\textmd{Im}\Sigma_2(\epsilon)}
{(x-\textmd{Re}\Sigma_2(\epsilon))^2+(\textmd{Im}\Sigma_2(\epsilon))^2}
\frac{\textmd{Im}\Sigma_1(\epsilon)}
{(x-E(0,\alpha)-\textmd{Re}\Sigma_1(\epsilon))^2+(\textmd{Im}\Sigma_1(\epsilon))^2}+\nonumber\right.\\
\left.
+\frac 12\sum_{n,\lambda}
\frac{\textmd{Im}\Sigma_1(\epsilon)}
{(x-E(n,\alpha)-\textmd{Re}\Sigma_1(\epsilon))^2+(\textmd{Im}\Sigma_1(\epsilon))^2}
\frac{\textmd{Im}\Sigma_1(\epsilon)}
{(x-E(n+1,\lambda)-\textmd{Re}\Sigma_1(\epsilon))^2+(\textmd{Im}\Sigma_1(\epsilon))^2}
\right]
\end{gather}
is the dimensionless conductivity kernel.
With the use of these, we obtain the various transport coefficients  as usual:
\begin{gather}
\sigma=\frac{2e^2}{\pi h}L_0,\\
S=\frac{1}{e}\frac{L_1}{L_0},\\
\frac{\kappa}{T}=\frac{2}{\pi h}\left(L_2-\frac{L_1^2}{L_0}\right),\\
L=\frac{\kappa}{\sigma T}=\frac{1}{e^2}\frac{L_2L_0-L_1^2}{L_0^2}.
\end{gather}
Here, $\sigma$ is the electric conductivity, $S$ is the Seebeck coefficient, $\kappa$ is the heat conductivity, where 
the last term ensures that the energy
current is evaluated under the condition of vanishing electric current, and $L$ 
is the Lorentz number. 
Off diagonal components of the conductivity tensors, such as the Nernst coefficient, are also of prime interest, but 
they cannot be simply evaluated from Kubo 
formula. Even in the case of a normal metal with parabolic dispersion, the Kubo formula turned out to be 
invalid\cite{girvin, oji}, and additional corrections have been worked out. Their determination for two-dimensional 
Dirac fermions is beyond 
the scope of the present investigation.

\begin{figure}[h!]
\psfrag{x}[t][b][1][0]{$\theta$ ($^\circ$)}
\psfrag{y}[b][t][1][0]{$\sigma\pi h/e^2$}
\centering{\includegraphics[width=7cm,height=7cm]{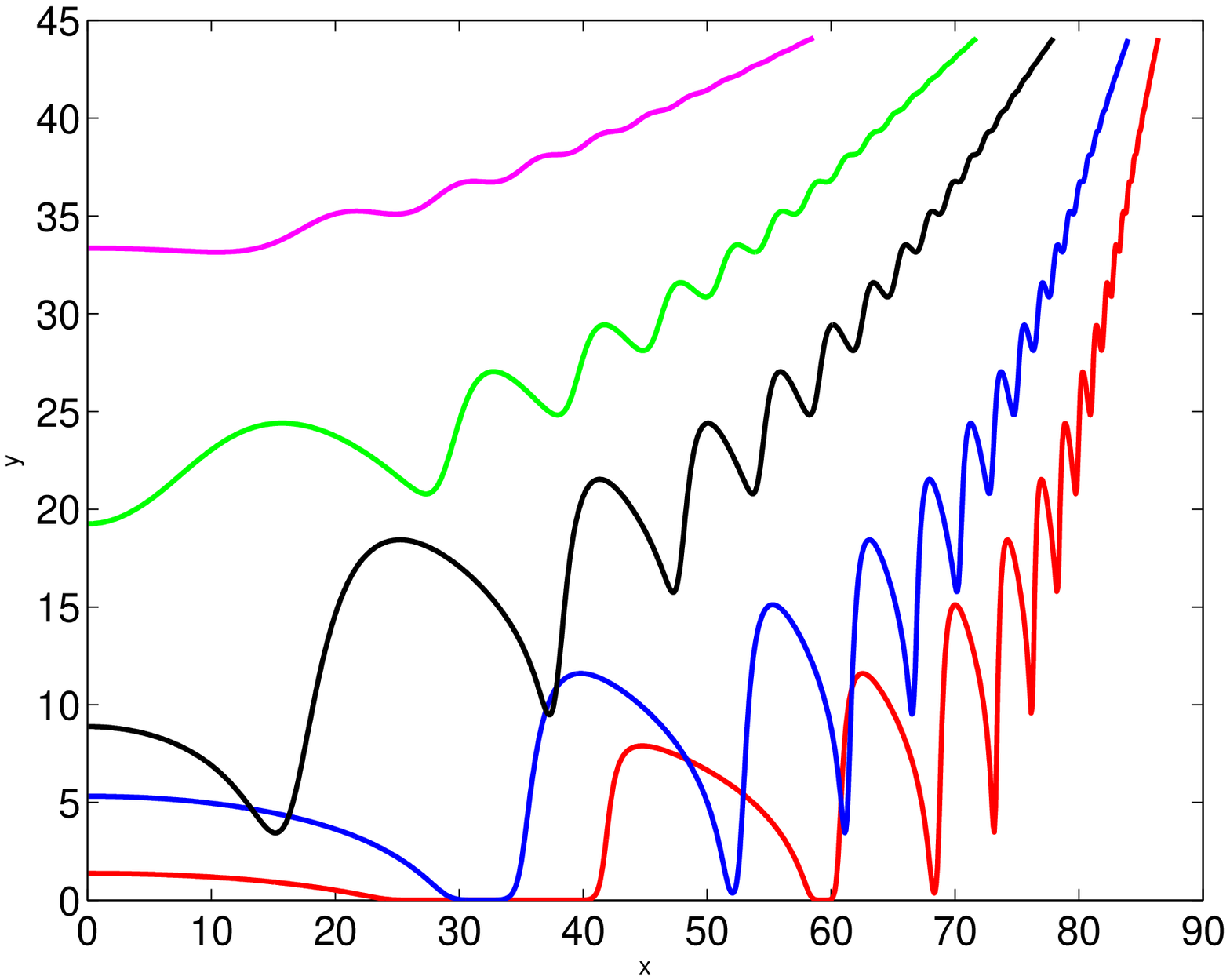}}
\hspace*{1cm}
\psfrag{x}[t][b][1][0]{$|B\cos(\theta)|/B_0$}
\psfrag{y}[b][t][1][0]{$\sigma\pi h/e^2$}
\centering{\includegraphics[width=7cm,height=7cm]{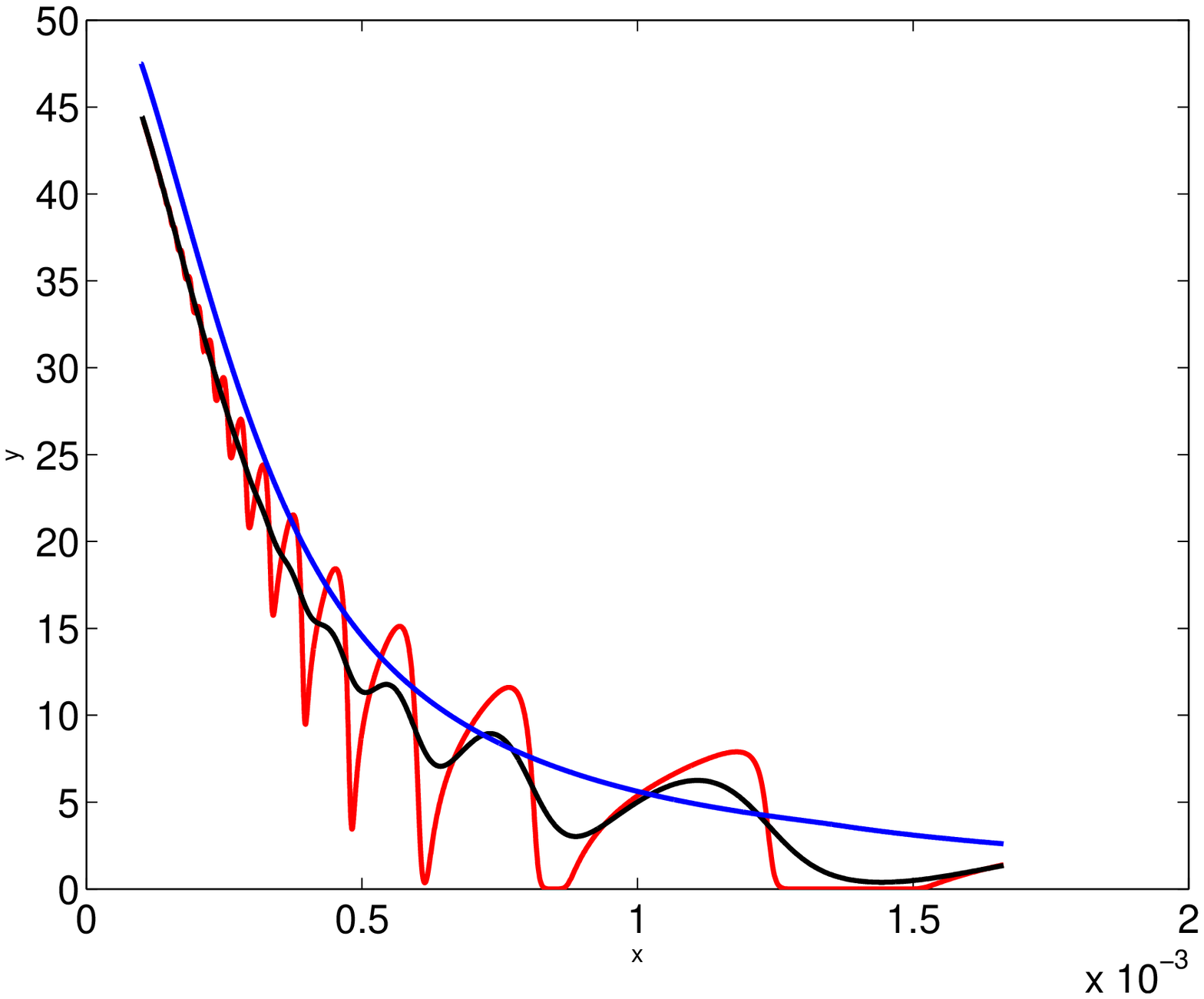}

\vspace*{-6.5cm}\hspace*{10cm}
\psfrag{x}[t][b][1][0]{$B_0/|B\cos(\theta)|$}
\psfrag{y}[b][t][1][0]{$\sigma\pi h/e^2$}
\includegraphics[width=4cm,height=4cm]{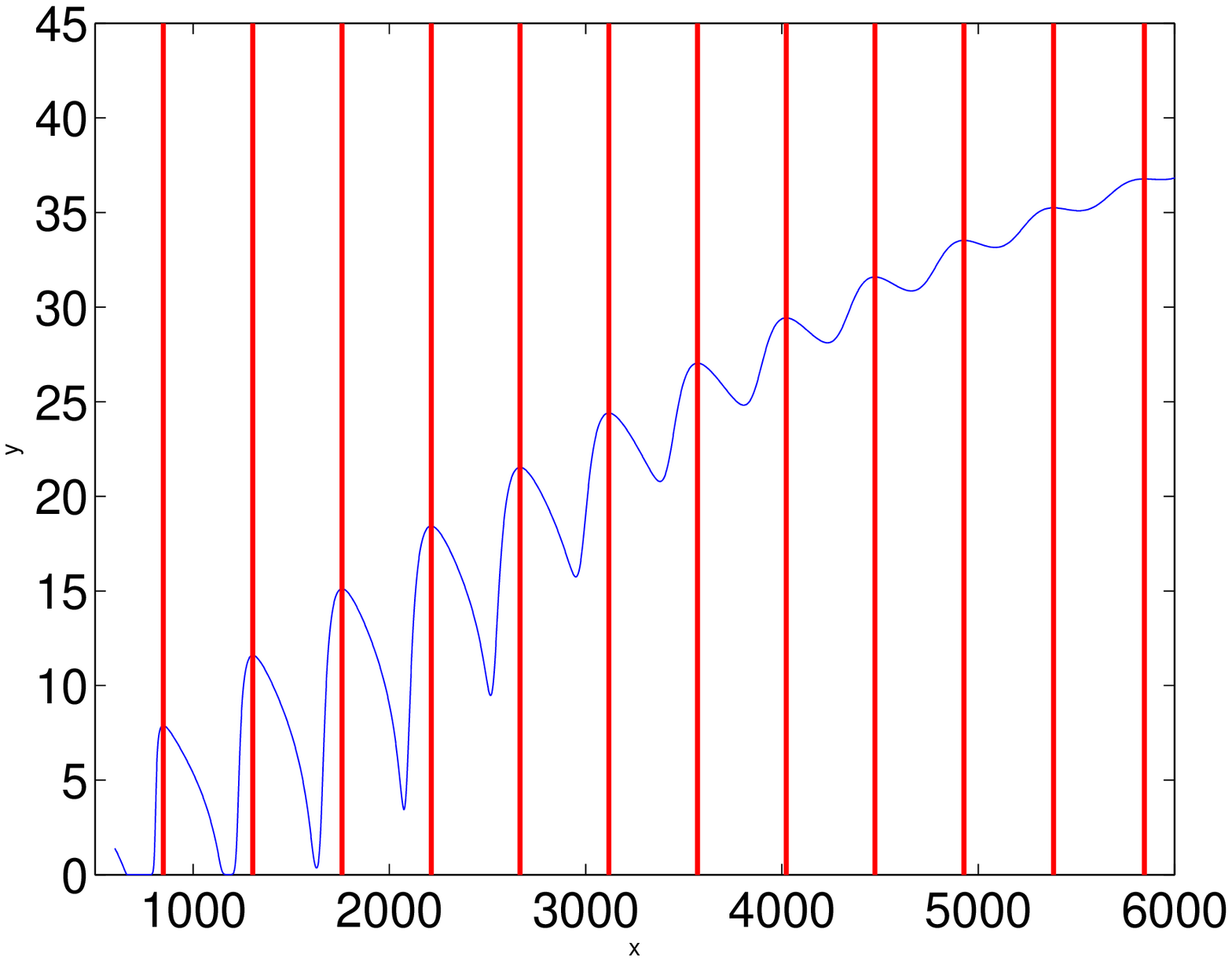}
\vspace*{3cm}}

\caption{(Color online) The angular dependent magnetoconductivity  oscillations are visualized for $\mu=0.05 D$,
$n_i$=0.001 and $T=0.0001D$, for magnetic fields $N=600$ (red), 1000 (blue), (2000) (black), (3000) (green) and 5000
(magenta) in the left panel. With increasing field (decreasing $N$), the oscillations become more pronounced, 
signaling
the discrete Landau level structure.
The right panel shows the electric conductivity for $\mu=0.05 D$,
$n_i$=0.001 and $T/D=0.0001$ (red), 0.001 (black) and 0.01 (blue).
For higher field, we arrive to the region, where crossing diagrams need to be
taken into account. The inset shows the electric conductivity as a function of $1/|B\cos(\theta)|$ to emphasize its 
periodicity.
\label{admr}}
\end{figure}

For the particle-hole symmetric case ($\mu=0$), the Seebeck coefficient is trivially zero.
If we consider the zero temperature, half-filled case, and assume small magnetic fields, we obtain the universal conductivity given 
by
\begin{equation}
\sigma_0=\frac{2 e^2}{\pi h},
\end{equation}
and similarly for the thermal conductivity as 
\begin{equation}
\frac{\kappa}{T}=\frac{2 e^2\pi}{3 h}.
\end{equation}
The Seebeck coefficient is zero. From this, the Lorentz number takes its universal value 
\begin{equation}
L_{u}=\frac{\pi^2}{3}\left(\frac{k_B}{e}\right)^2,
\end{equation}
which means, that in this limit, the Wiedemann-Franz law holds\cite{klasszikus,sharapov1}.
Landau levels always develop around the meeting point of the conical valence and conduction band. If we are at half 
filling ($\mu=0$), no levels cross 
$\mu$ when varying the magnetic field, since they are symmetrically placed below and above. However, when $\mu$ is 
finite, Landau levels can cross its value with changing the field, and we expect Shubnikov-de Haas oscillations. In 
general, when the number of levels below $\mu$ is 
large (or $\omega_c\ll |\mu|$), we can conjecture the periodicity of these oscillations. Assume that a level (the 
$n$th) sits right 
at the chemical potential ($\mu=\omega_c\sqrt{n+1}$). Then, the distance from the adjacent level determines the period 
of the oscillations. This is 
\begin{gather}
|E(n+1,\alpha)-E(n,\alpha)|\approx 
\frac{\omega_c}{2\sqrt{n+1}}=\frac{\omega_c^2}{2\mu}=\frac{v_F^2e|B\cos(\theta)|}{\mu}\sim B
\label{szinttav}
\end{gather}
provided, that $n\gg 1$. This means, that albeit the Landau levels show an unusual $\propto\sqrt{n}$ dependence of the 
level 
index compared to that in a normal metal $\propto n$, the Shubnikov-de Haas oscillations turn our to be still periodic 
as a function of $1/B$. 
The comparison of the coefficient of the magnetic field in Eq. \eqref{szinttav} to that in a parabolic 
band\cite{abrikosov} suggests, that the cyclotron mass can be defined as $m_c=\mu/v_F^2$. Even though the spectrum is 
linear, the finite chemical potential provides us with an energy scale for $m_c$\cite{novoselov2}. 
This can readily be checked in Fig. \ref{admr}, where not only the field, but the angle 
dependence of the conductivity of shown for different field strength.  The larger the magnetic field, the more visible 
the oscillations are, although these can be smeared by increasing the concentrations.

The explicit value of the chemical potential, which is fixed by the particle number at a given temperature and
field,  should also be determined self-consistently. However, no serious deviations from its initial values have been 
detected during the evaluation process, and these did not affect the dependence of physical quantities on $T$ and $B$ 
in the investigated range of parameters. Presumably, taking a large value of the chemical potential would require 
its self-consistent determination as well.

\begin{figure}[h!]

\psfrag{x}[t][b][1][0]{$B_0/|B\cos(\theta)|$}
\psfrag{y}[b][t][1][0]{$L/L_u$}
{\includegraphics[width=7cm,height=7cm]{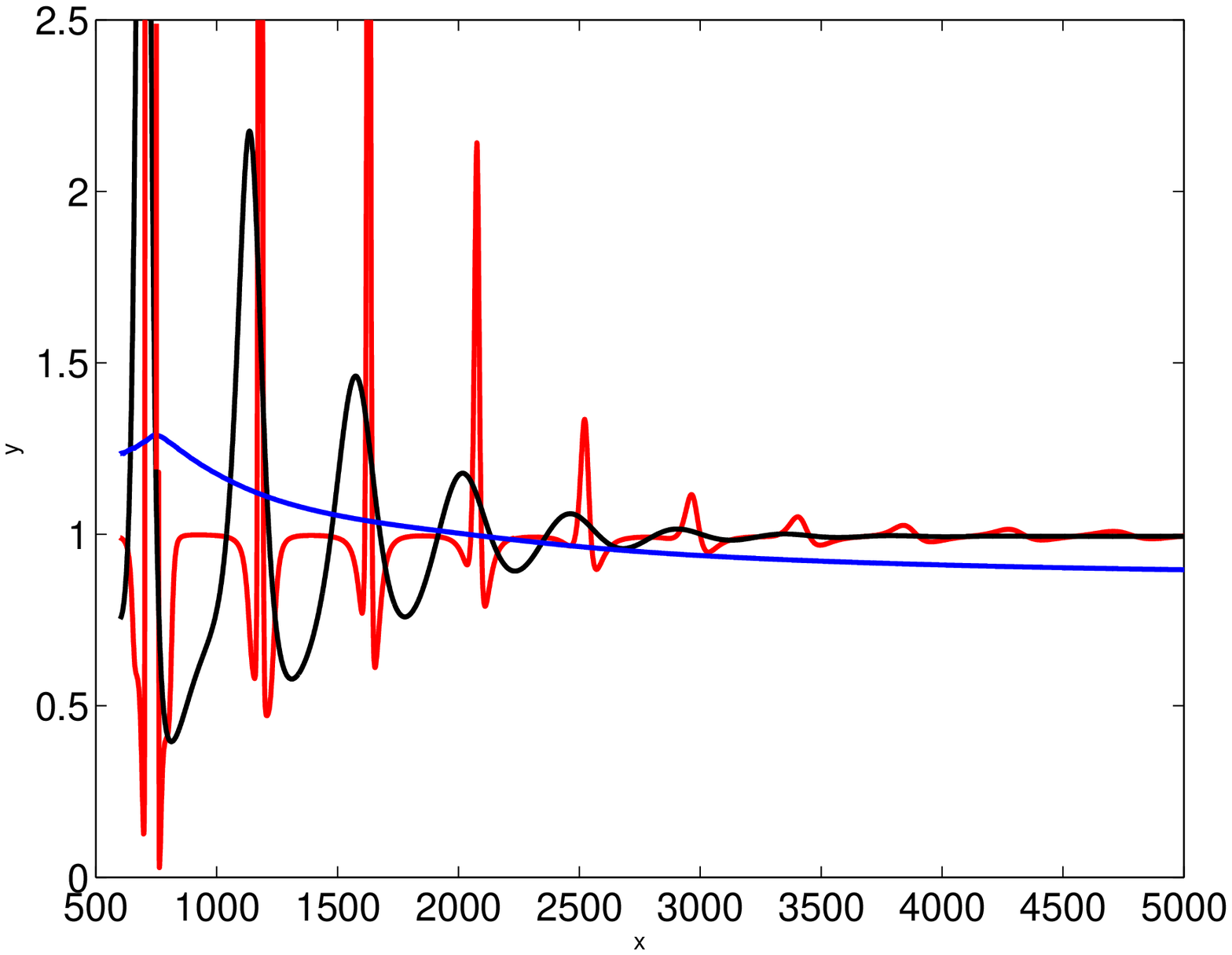}}
\hspace*{10mm}
\psfrag{x}[t][b][1][0]{$|B\cos(\theta)|/B_0$}
\psfrag{y}[b][t][1][0]{$S e$}
\psfrag{y}[b][t][1][0]{$\kappa\pi h/T$}
{\includegraphics[width=7cm,height=7cm]{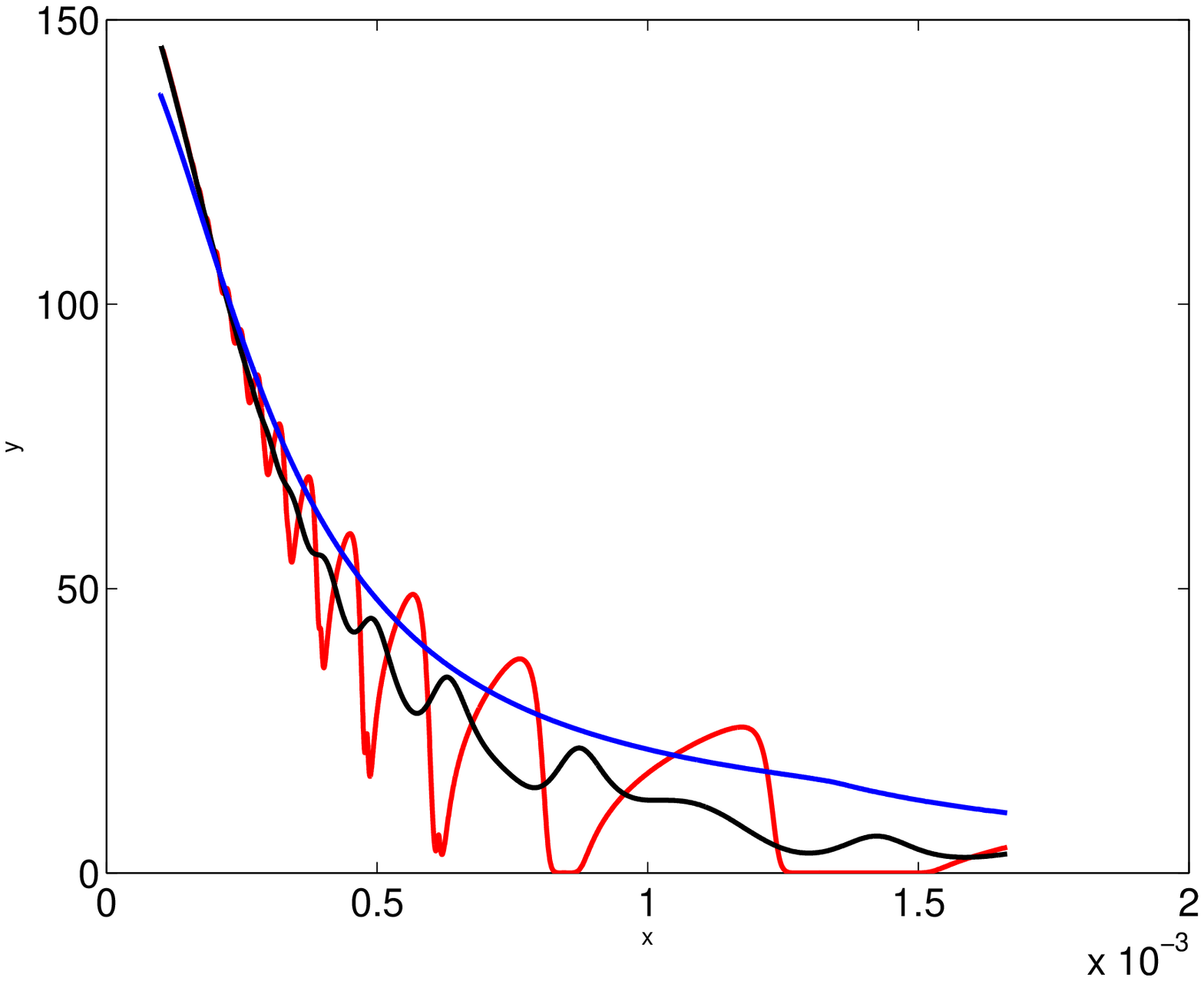}}

\vspace*{-6.7cm}\hspace*{10.3cm}
\psfrag{y}[b][t][1][0]{$S e$}
{\includegraphics[width=4cm,height=4cm]{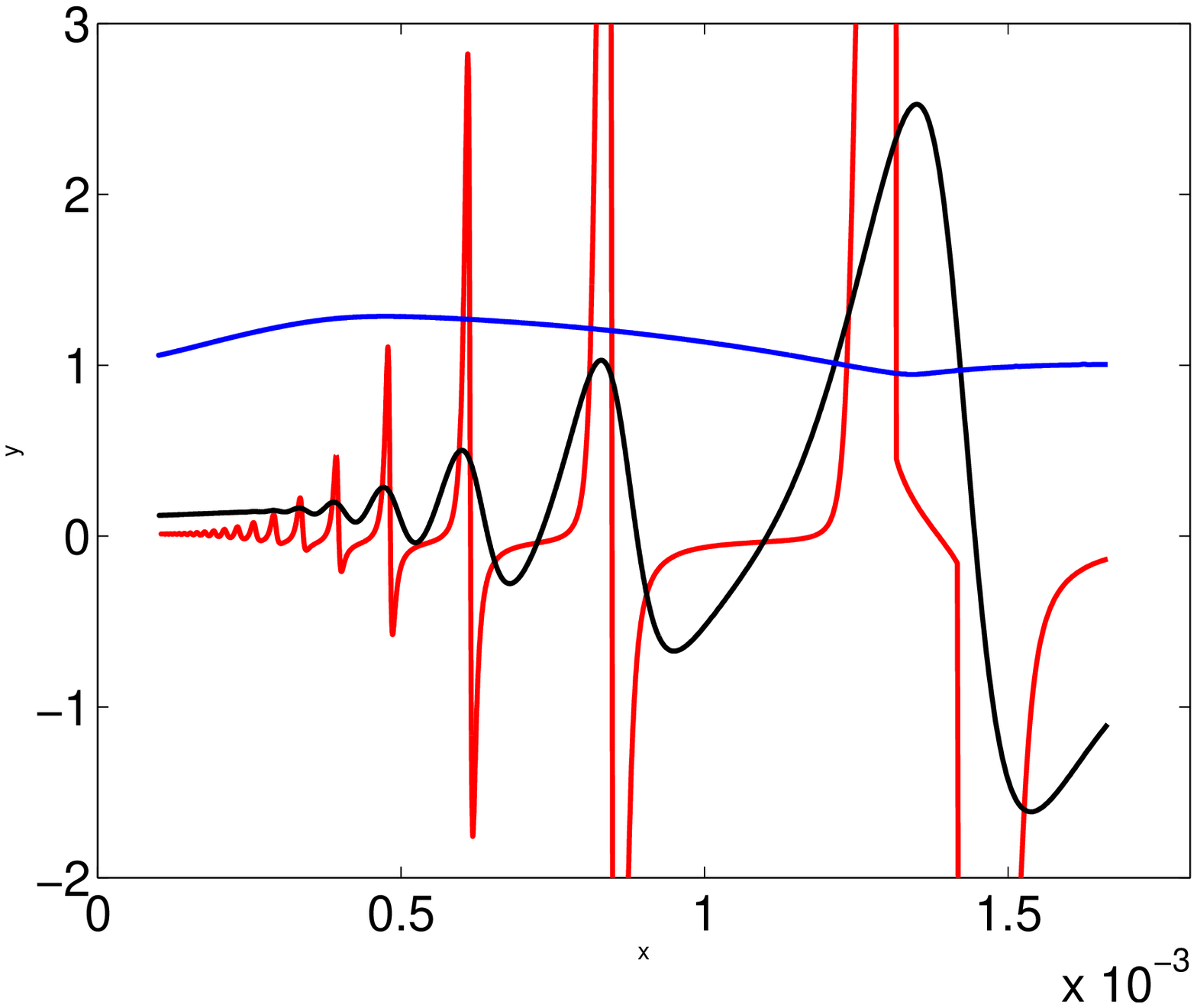}}
\vspace*{3cm}

\caption{(Color online) The left panel shows the Lorentz number as a function of the inverse magnetic field to sress the 
periodic violation of the Wiedemann-Franz law for $\mu=0.05 D$,
$n_i$=0.001 and $T/D=0.0001$ (red), 0.001 (black) and 0.01 (blue). The right panel shows the 
heat conductivity and the Seebeck coefficient (inset) for the set of same parameters. \label{kappaB}}
\end{figure}

In Fig. \ref{kappaB}, we show the magnetic field dependence of the heat conductivity. It resembles closely to the 
electric one at low temperatures. However, at higher temperatures, each peak in the oscillations split into two. This 
occurs, because in the electric conductivity, the kernel is sampled by the $1/\cosh^2((\epsilon-\mu)/2T)$ function, 
which 
gathers 
information about excitations at the chemical potential. However, an extra $(\epsilon-\mu)^2$ factor appears in the 
heat response, which measures the immediate vicinity of $\mu$ above and below, within a window $2T$, which gives the 
splitting.
The oscillations become smoothened with decreasing field, in contrast to Ref. \onlinecite{sharapov3}, where large 
oscillations 
were found even at small fields. The difference can be traced back to our field dependent scattering rate 
(Eqs. \eqref{self1}-\eqref{self2}), as opposed to the field independent one used in Ref. \onlinecite{sharapov3}.
Similar features have been observed in highly oriented pyrolytic graphite\cite{ocana,ulrich}.
By decreasing the field, $N$ increases, and the density of states becomes similar to that of a d-wave 
superconductor\cite{hottacomment}, without significant deviations from linearity.
Both $\sigma$ and $\kappa$ decrease with field, a feature already present at $\mu=0$. As we increase the field, 
$\omega_c$ increases, and so does the distance between Landau levels. Then, at a given temperature, a smaller  
number of 
states will be present for excitations around $\mu$, hence the corresponding conductivity decreases. 
The Seebeck  coefficient 
shows sharp oscillations which die out with temperature. Its background value, after subtracting the oscillations, 
is found to be almost magnetic field independent, but smoothly increases with temperature.
The Lorentz number remains close to one, if we subtract the oscillations. However, due to the double (single) peak 
structures in the heat (electric) response, their ratio shows wild, but sharp deviations from unity at specific fields, 
where the 
Wiedemann-Franz law is violated. In contrast to this, one would have encountered large and wide oscillations in the 
Lorentz number as a function of field in the presence of phenomenological, constant scattering rate.

\begin{figure}[h!]

\psfrag{x}[t][b][1][0]{$\mu/D$}
\psfrag{y}[b][t][1][0]{$\sigma\pi h/e^2, \kappa 3h/T\pi$}
{\includegraphics[width=7cm,height=7cm]{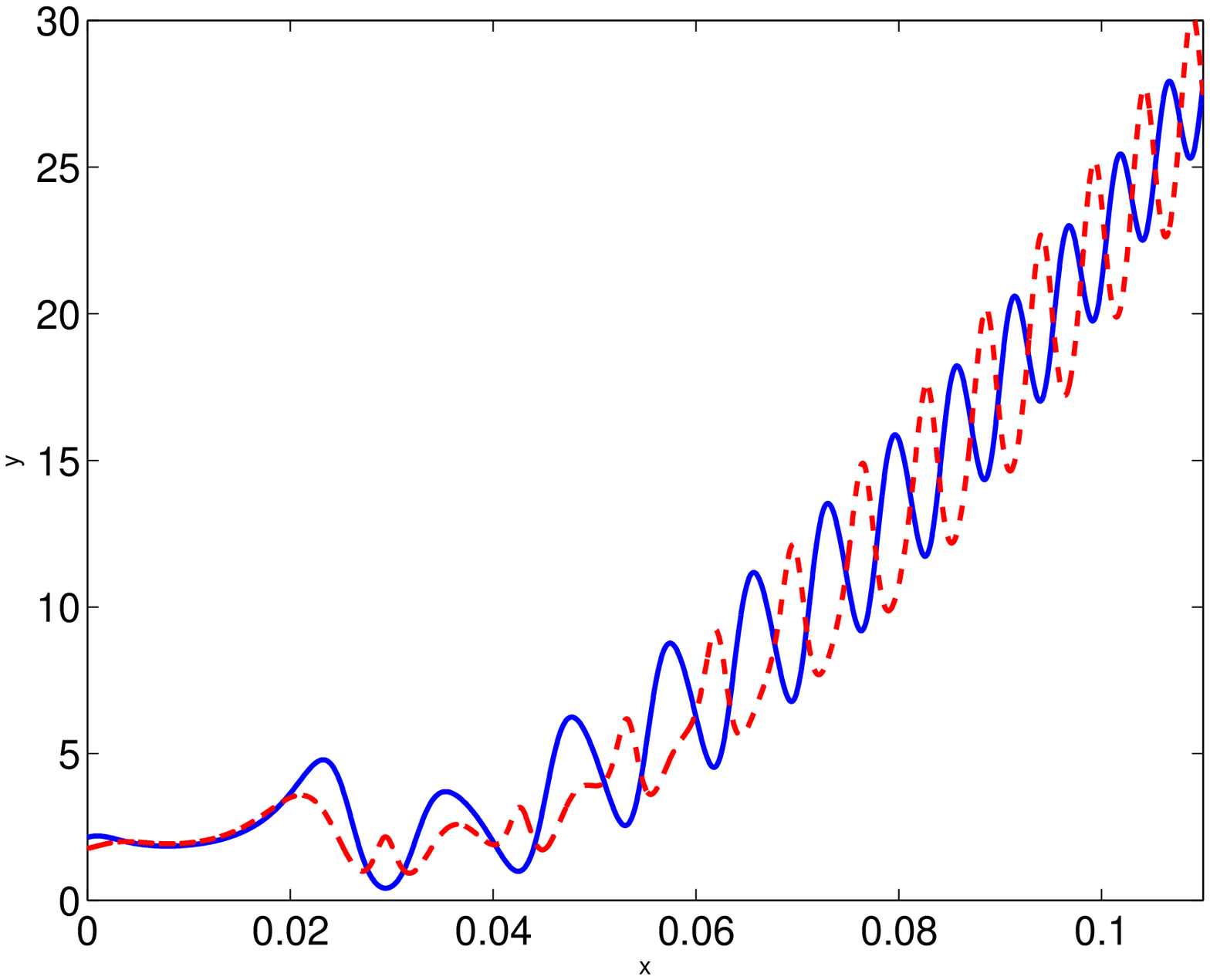}}

\vspace*{-6.6cm}\hspace*{-2cm}
\psfrag{y}[b][t][1][0]{$Se$}
{\includegraphics[width=3.5cm,height=3.5cm]{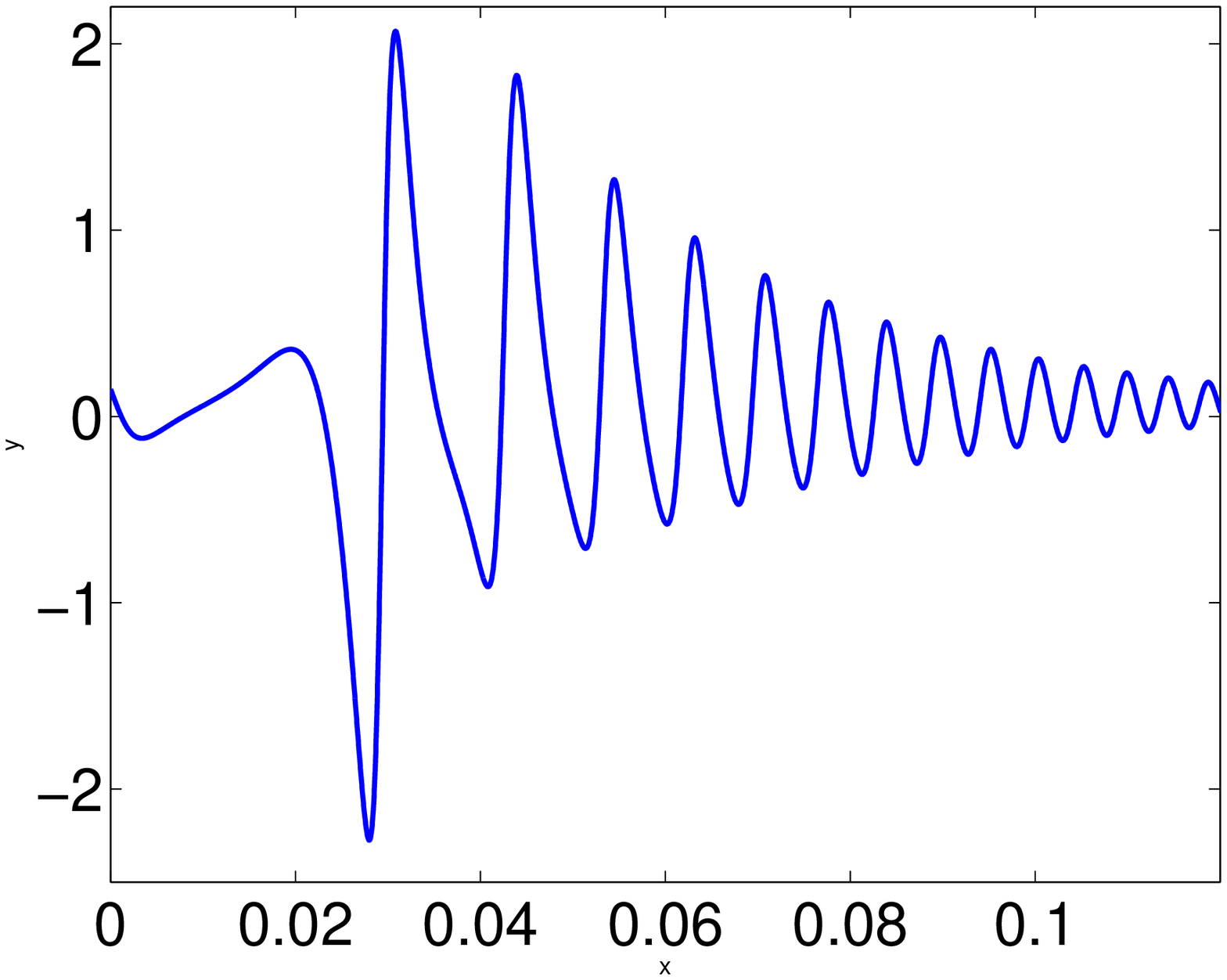}}
\vspace*{3cm}

\caption{The electric (blue solid line) and heat (red dashed line) conductivity and the Seebeck coefficient (inset) are 
shown as a function of the chemical 
potential for $T=0.001D$, $N=1000$, $n_i=0.001$. Due to the antiphase oscillations, the 
Wiedemann-Franz law is violated.\label{sigmavsmu}}
\end{figure}

In Fig. \ref{sigmavsmu}, we show the evolution of the electric and heat conductivity and the Seebeck coefficient as a 
function
of chemical potential. In accordance with experiment in Ref. \onlinecite{novoselov2}, we also find oscillations, 
corresponding to Landau levels, which also smoothen with temperature. 
Interestingly, the splitting of the peaks in the heat conductivity is nicely observable as a function of $\mu$. These 
occur in such a way that they produce antiphase oscillations with respect to the electric one, and lead to the violation 
of the Wiedemann-Franz law.
The Seebeck coefficient shows peculiar behaviour. 
At the particle-hole symmetric case, it is zero, and remains mainly so apart from large oscillations.

\begin{figure}[h!]

\psfrag{x}[t][b][1][0]{$T/D$}
\psfrag{y}[b][t][1][0]{$\sigma\pi h/e^2, \kappa h/T\pi$}
{\includegraphics[width=7cm,height=7cm]{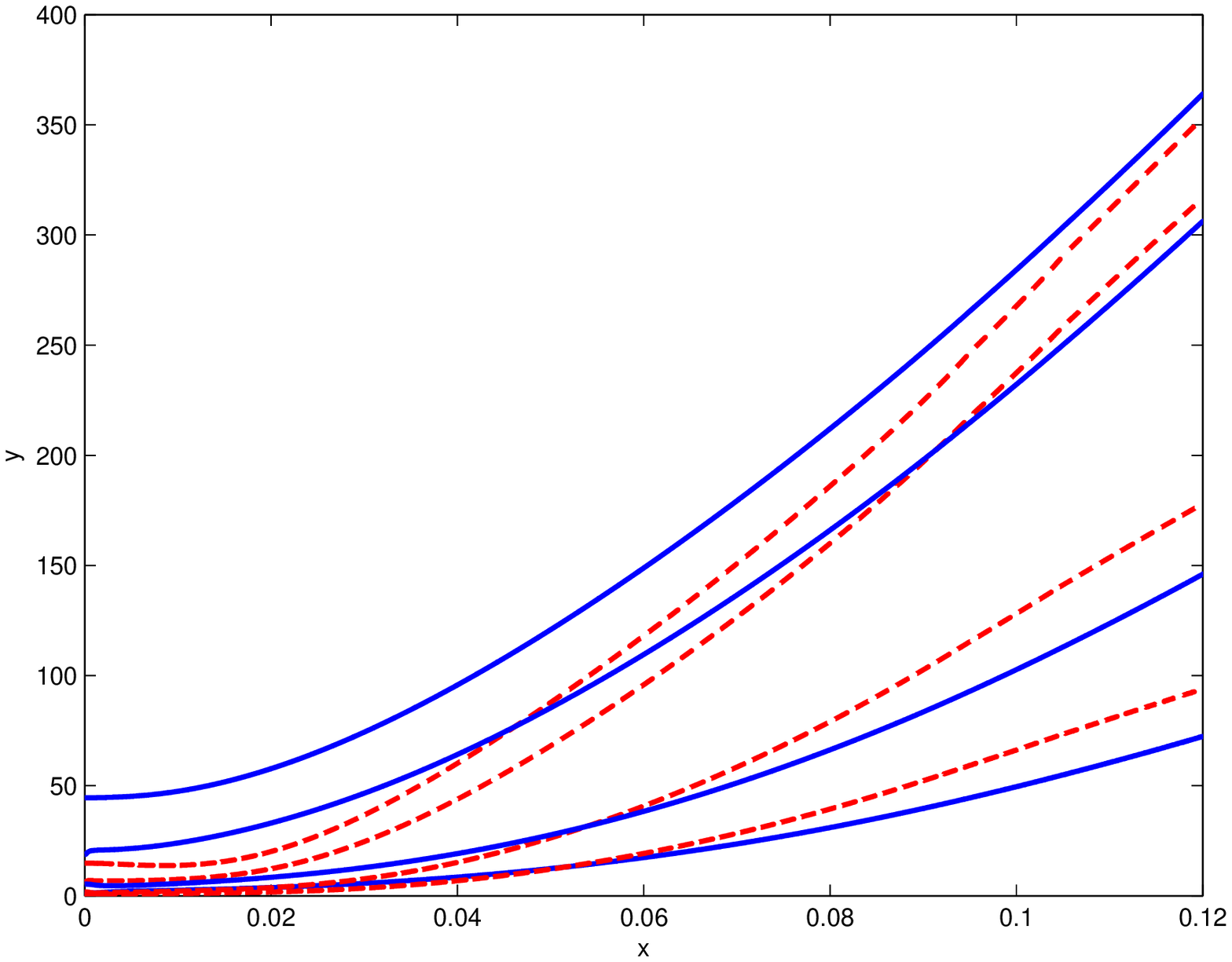}}
\hspace*{7mm}
\psfrag{x}[t][b][1][0]{$T/D$}
\psfrag{y}[b][t][1][0]{$S e$}
\centering{\includegraphics[width=7cm,height=7cm]{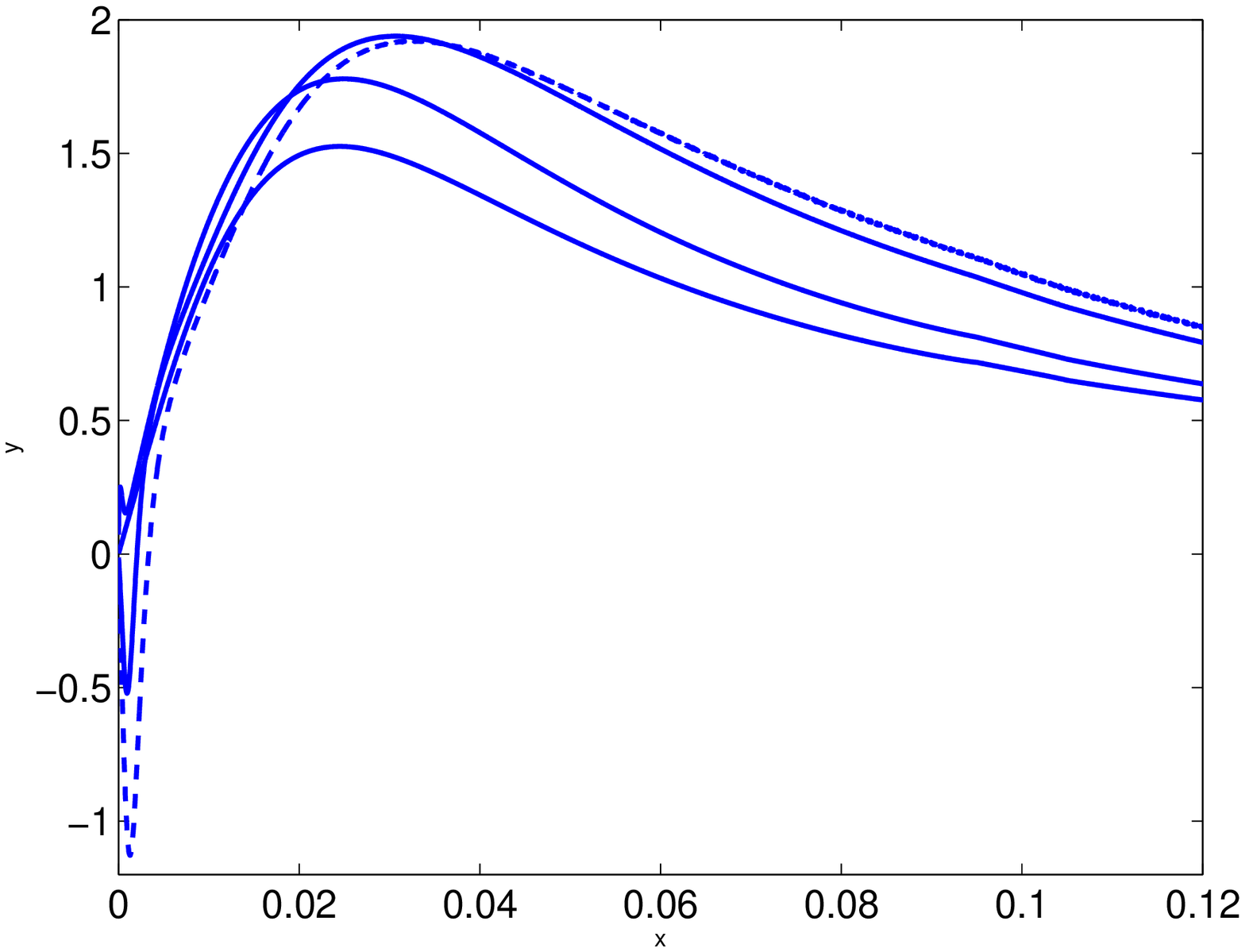}}

\vspace*{-4.3cm}\hspace*{7.3cm}
\psfrag{y}[b][t][1][0]{$L/L_u$}
\centering{\includegraphics[width=4cm,height=3.3cm]{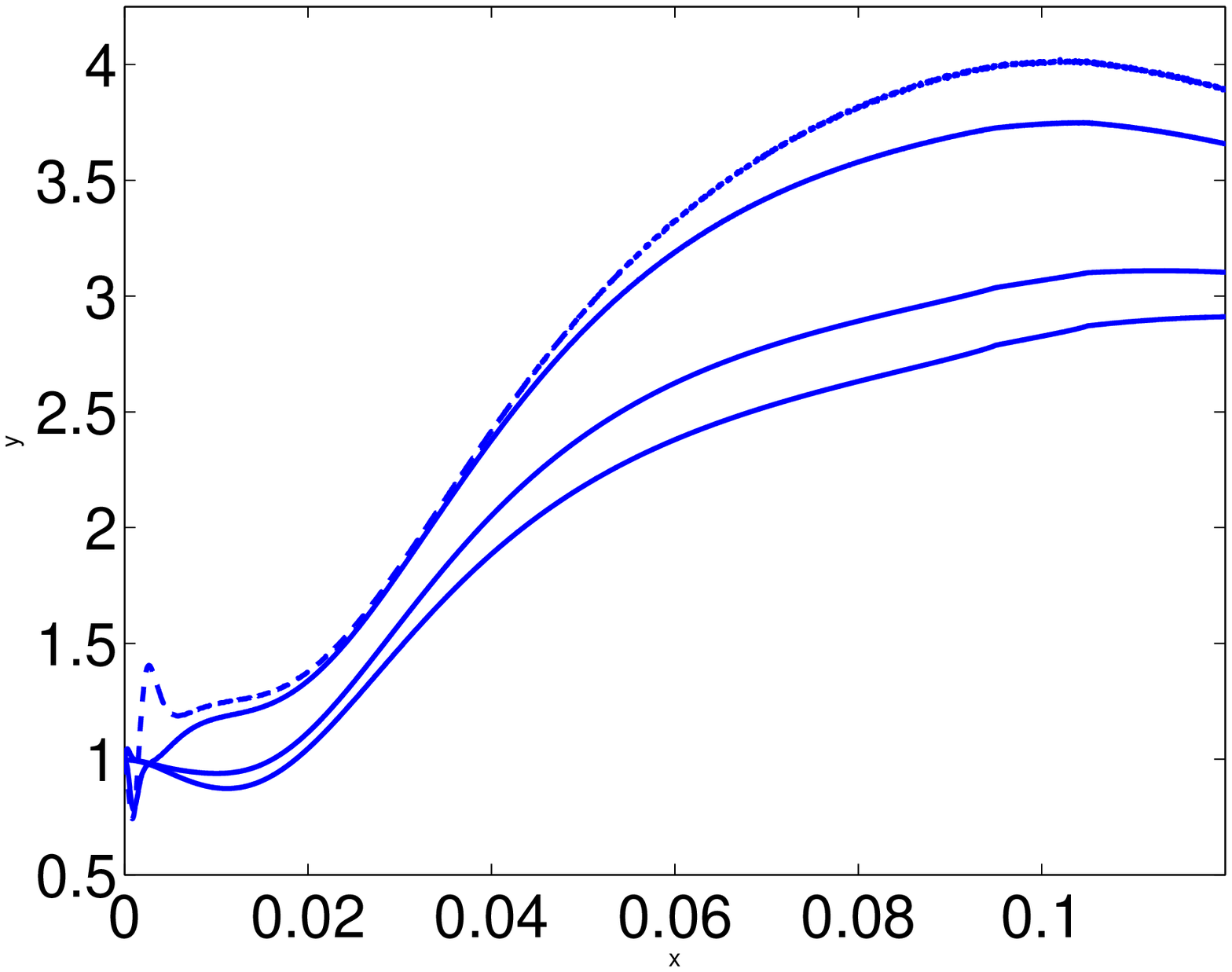}}
\vspace*{1cm}

\caption{(Color online) The electric (blue solid line) and heat (red dashed line) conductivity  
are shown in the right panel for $n_i$=0.001, $\mu=0.05D$, $N$=600, 1000, 3000 and 10000 from bottom to top.
Note the $1/\pi^2$ reduction of the heat conductivity.
The right panel shows the Seebeck coefficient and the Lorentz number (inset) for the same parameters
from top to bottom, with dashed line for $N=600$. Note the violation of the Wiedemann-Franz law at low 
temperatures at high fields (smaller $N$)!\label{kappaT}}
\end{figure}

The temperature dependence of the electric and heat response is shown in Fig. 
\ref{kappaT}. Both increase steadily with temperature, since more available states are accessible with $T$. However, at 
small temperatures, a small decrease is observable in low fields, in accordance with other 
studies\cite{peresalap,sharapov1}
The Seebeck coefficient first increases, and after a broad bump, decreases with $T$. For higher temperatures, the 
bandwidth $D$ makes its presence felt. The Wiedemann-Franz law remains intact at low temperatures and fields, but 
becomes violated for higher $T$ or $B$.

\section{Conclusion}

We have studied the effect of localized impurities in two dimensional Dirac fermions in the presence of quantizing, 
arbitrarily oriented magnetic field. 
The energy spectrum depends on the level index as $\propto\sqrt{n}$, as opposed to the $n+1/2$ linear dependence in 
normal metals\cite{klasszikus}.
Expressions for both the electric and heat current in the presence of magnetic field were worked out.
The self-energy in the full Born-approximation obeys self-consistency conditions, resulting in important magnetic field 
and frequency dependence of scattering rate and level shift. In the density of states, only a small island shows up 
close to zero frequency for small 
fields, similarly to d-wave superconductors\cite{impurd-wave}. By increasing the field, oscillations become visible, 
corresponding to Landau levels. By further increasing the field, these become separated from each other, and clean gaps 
appear between the levels, in which intragap states, small islands show up at high field. 
The non-Lorentzian broadening of Landau levels and the intragap features differ from previous studies assuming a 
constant 
scattering rate, and should be detected experimentally in graphene.

Both the electric and thermal conductance shows Shubnikov-de Haas oscillation in magnetic field, which disappear for 
small fields and higher temperatures. These are periodic in $1/B$, similarly to normal metals, in spite of the 
different 
Landau quantization.
The Seebeck coefficient shares these features, but its oscillations are really large as opposed to $\sigma$ and 
$\kappa$. The Wiedemann-Franz law stays close to unity, except at certain fields, where large deviations are 
encountered, 
which vanish with decreasing field.
Besides oscillations, both $\sigma$ and $\kappa$ decreases with field, since the larger the cyclotron frequency, the 
smaller the probability of finding states around $\mu$.
These are in agreement with experiments on the thermal conductivity of highly oriented pyrolytic 
graphite\cite{ocana,ulrich}.
Oscillations are also present as a function of chemical potential, similarly to experimental findings\cite{novoselov3}.

The temperature dependence of the conductivities is rather conventional, both $\sigma$ and $\kappa$ increases with 
temperature steadily, regardless to the value of the chemical potential. The Seebeck coefficient exhibits a broad bump 
around $T\sim \mu$, and decreases afterwards. The Wiedemann-Franz law is obeyed for small temperatures and field, but 
violated for higher values.

\begin{acknowledgments}
This work was supported by the Hungarian  
Scientific Research Fund under grant number OTKA TS049881. 
\end{acknowledgments}

\bibliographystyle{apsrev}
\bibliography{refgraph}
\end{document}